# PSN: Portfolio Social Network

Jordi M. Cortes, Sarwat Nizamani and Nasrullah Memon

*Abstract*—In this paper we present a web-based information system which is a portfolio social network (PSN) that provides solutions to recruiters and job seekers. The proposed system enables users to create portfolios so that he/she can add his specializations with piece of code, if any, specifically for software engineers, which is accessible online. The unique feature of the system is to enable the recruiters to quickly view the prominent skills of the users. A comparative analysis of the proposed system with the state of the art systems is presented. The comparative study reveals that the proposed system has advanced functionalities.

## I. INTRODUCTION

For a student after completion of graduation, the next critical phase is to find a job that best suits his/ her educational level and interests. A minimum criterion is set by the employer for the job position which includes educational, experience, eager to learn, etc. The recruiter then chooses a method to advertise the position in order to publicize the required positions.

One of the widely used methods is newspaper advertisement [1]. The local employment office such as the Job Centre is another source where the recruiter can find the employees. The modern age of Internet also has affected the process of advertisement and recruitment. Searching for jobs on the internet can be benefited from making his/ brand [2]. The professionals of today use more often social media than reading the newspaper [1]. The social network sites (SNS) let the user create a public or partially public profile that can be seen by other users [3]. SNS also gives suggestions to users to make new connections. SNS allow users to make social relationships, share the interests, ideas, events, experiences, etc. There are some social networking sites [3] which are specifically used by professionals, recruiters and job seekers. These sites include LinkedIn [4], Viadeo [5] and Xing [6]. The existing systems other than social networks which facilitate job seekers, professionals and recruiters include VisualCV [7]. It enables users to create an online CV, build and manage an online career portfolio, and share professional qualifications with employers, professionals, partners and customers. LinkedIn is one of the largest social networking sites specifically used for professional networking purpose. It has more than 200 million users in more than 200 countries. It is used for the purpose of finding job according to one's professional skills, recruiting perfect candidate for required position, acquiring help from professionals in one's field of interest.

The motivation of the PSN is highlighted in the following comment by a recruiter: " *When I've interviewed developers, they rarely bring any samples of their work to show*" [8]. The comment motivated us to design such a framework which enable users not to tell their skills but show them.

PSN is a career portfolio that enables users to organize personal information, qualifications and professional skills. Users not only can write about their skills but also can show their work samples. In addition, it also has the functionalities of the professional social networks.

The main idea behind the development of PSN was to enhance visibility of the skilled applicants, by showing their work to others not only telling.

The key benefits of the PSN in addition to existing professional social networks include:

(i) it enables applicants to show their skill to others;

(ii) applicants can make their interests, eagerness to learn and motivation visible to others; otherwise these would have been unnoticed;

(iii) PSN helps recruiters in making decisions to select competent and suitable candidates for required positions, because all the skills, experience, and interests of the candidates are visible to them.

 The rest of the paper is organized as follows: Section 2 discusses related work in connection to proposed system, while Section 3 describes the proposed system. Section 4 illustrates the results and Section 5 concludes the paper.

## II. RELATED WORK

This section presents the related work in connection to PSN. The most closely related system to the proposed system are existing professional social network sites such as LinkedIn, Viadeo and Xing. These networks provide functionalities comparable to PSN and are considered to be the state of the art . The comparison table is given in Section 4, which reveals that the way PSN is more viable system than the existing systems.

LinkedIn is the most widely used social networking site by professionals since 2003 [2]. LinkedIn is considered to be a business-oriented social network. The users typically are affiliated with social groups of their professional interests [9]. It is also being used for finding jobs, potential candidates for job position and business opportunities. According to research [10], the most often used social networking site by organizations for recruitment is LinkedIn, which is 95% of all the social networking sites for recruitment purpose. In comparison to PSN, LinkedIn does not have functionality for adding information on collaborating projects carried out by the user, and no availability for adding code snippet.

Viadeo connects more than 50 million professionals. Unlike PSN, Viadeo has no concept of portfolio, no availability of creating projects and adding code snippet.

Xing is one more business oriented social network, especially focused in Europe [11]. It is a medium sized social network with more than 11 million users. It was launched as business social networking site in October 2003 [12]. Like Viadeo, Xing also does not provide functionality of adding projects and snippet and does not consider the concept of portfolio.

VisualCV is another system that does contain the concept of portfolio; however, it is not social networking site. It also does not allow candidates to create projects, nor allows candidates to add project collaboration information, though it allows candidates to add the code as image.

One article [12] focuses on revenue model of various social networking sites. Three aspects of social networking sites which contribute to revenue generation are discussed. These include advertising models; subscription models; and transaction models. A comparative study of various social networking sites is presented specifically in revenue generation.

## III.  PROPOSED SYSTEM

The proposed system is elaborated by first describing architecture followed by implementation details.

PSN has been designed under Model-View-Controller (MVC) architecture [13] which is proved to be a good solution for web-based applications. The architecture is depicted in Fig 1 as described in one article [14]. The Model layer of the architecture is responsible for managing the behavior and data of the application, responding to requests for information about its state and instructions to change its state [14]. The View layer is responsible for displaying the information. The Controller layer is responsible for user interface to inform the model or view layer about user requirements.

Typically, MVC architecture works as follows:
- The browser sends request to the server that hosts the MVC application
- The controller is invoked to handle the request
- The controller interacts with the model, possibly resulting in a change in the model's state
- The controller invokes the view
- The view renders the data (often as HTML) and returns it to the browser to display

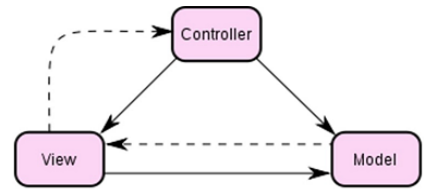

Fig.1. MVC architecture

The logic design of PSN is implemented in PHP, which is extensively used for dynamic web applications development. The PHP code can easily be embedded in an HTML source document and processed by a web server with PHP processing module in order to generate a web page.

The use of a framework for development of a web application alleviates the work of developers by making easier the most common tasks, such as database access, session management for user or template management. The framework used for the PSN is Yii [15], which is one of the high performance PHP framework for Web 2.0 applications and supports MVC-based architecture. It is fast because it only loads those features of application which are needed. It provides great caching support and is a secure framework having security as standard.

The database has been designed using MySQL, which is open source relational database, proved to be a scalable solution for web applications. Its phpMyAdmin tool enables easy handling of database administration. The database for the PSN can be described in Fig.2.

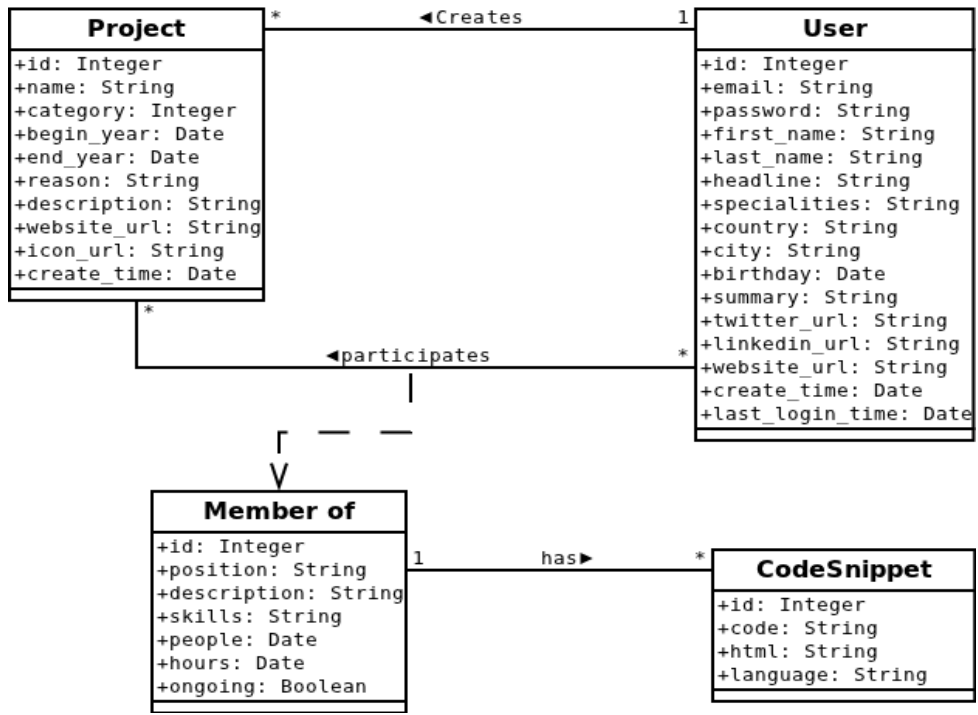

Fig. 2. Conceptual Model of the

After describing architectural, implementation and database detail, we now discuss the main functionalities of the PSN.

These include:

- Creation and registration of user
- Adding and updating user information
- Creating and updating project
- Creating and updating a member
- Describing each member's responsibility in project
- Adding code snippet, in which user can add codes to show his/ skills
- PSN is a web application that can be accessed by most of the web browsers
- Reliability and efficiency are considered critical requirements for online applications, the use of Yii framework and MVC architecture make this possible
- Maintenance and security are also under consideration; the framework used allows easy management of security
- Scalability is another issue for web applications; a MySQL database is used to satisfy this requirement

## IV. RESULTS

In this article we presented the proposed system, i.e. a portfolio social network, which provides more functionalities than the existing systems. A comparative study has also been conducted. Table 1 illustrates the results of the comparative study. It can be observed clearly that PSN offers more functionality to enhance one's visibility in the professional social network. As it can be seen that PSN supports the concept of portfolios like LinkedIn, but unlike LinkedIn, PSN also allows users to add code snippet and project collaboration information. It can also be noted from the table that the other two systems, Viadeo and Xing, do not support the concept of portfolio and do not allow the creation of projects, thus no information on project collaboration. VisualCV supports the concept of portfolio but it does not come under the category of social networks. It does not allow users to create projects, therefore has no information on project collaboration. However, it allows users to add code as image.

The screenshot of the sample portfolio for a user is illustrated in Fig 3, demonstrating the work and skills of the user. One can see the projects carried out by the user along with the responsibilities of the user. The recruiters can see straightforwardly what potential skills a candidate has, which makes the recruitment process easier.

Table 1. Results comparing PSN to other platforms

| Features | PSN | LinkedIn | Viadeo | Xing | visualCV |
|---|---|---|---|---|---|
| **Concept of Portfolio** | Yes | Yes | No | No | Yes |
| **User Information** | Yes | Yes | Yes | Yes | Yes |
| Personal | Yes | Yes | Yes | Yes | Yes |
| Educational | No | Yes | Yes | Yes | Yes |
| Professional | Some | Yes | Yes | Yes | Yes |
| Presence in other social networks | Yes | Yes | Yes | Yes | Yes |
| **Creation of Project** | Yes | Yes | No | No | No |
| Information about | Yes | Yes | No | No | No |
| People in the project | Yes | Yes | No | No | No |
| **Project collaboration information** | Yes | No | No | No | No |
| Skills required | Yes | No | No | No | No |
| People in charge | Yes | No | No | No | No |
| Dedicated hours | Yes | No | No | No | No |
| Code Snippets | Yes | No | No | No | Yes* |
| Work images | No | No | No | No | Yes |
| People who you worked with | Yes | Yes | No | No | No |

\* Project code only in image

Fig. 3. Screenshot of a sample portfolio from PSN

## V. CONCLUSION

The paper presents the proposed system PSN, which is a portfolio social network. PSN has been developed as an aid to job seekers and recruiters in addition to professional social networks. PSN lets the user add real work as code snippet, which helps recruiters judge the development skills of the candidate. PSN is designed under modern technologies as a web application that also satisfies the needs of online systems such as efficiency, maintenance, security, scalability, etc. In the paper, a comparative study of existing systems with the proposed system is also presented, depicting the unique functionalities of the prospective system. In future, we aim to add a 'recommendation' feature so that users can recommend to each other the work of others. We also plan to add feedback functionality that will help us improve the PSN. Another future task we are considering is the implementation of role-based access control to assign roles to users in order to improve management of project permission requests.

**Jordi Magriña Cortes** studied Computer Software at the Polytechnic University of Catalonia and spent a semester abroad as an exchange student in the University of Southern Denmark.

At the present time he is developing applications for iOS and Android in a Barcelona-based consulting IT company. More specifically, his tasks are focused in the project management, implementation of the required functionalities and user interfaces.

**Sarwat Nizamani** received her B.Sc. (Hons.) and M.Sc. (Hons.) in Computer Science in the years 1998 and 1999 respectively from University of Sindh, Pakistan. The author received her Master's of Science in Robotic Engineering, from University of Southern, Denmark in 2011.

She worked as Research Associate from 2000-2003, then as lecturer from 2003-2007 and as Assistant Professor from 2007 to-date in University of Sindh, Pakistan. Since April 2010, she is PhD student, at University of Southern, Denmark. She has total eight publications in journals/ conferences/ book chapters of national and international repute. Her field of research interest is machine learning, natural language processing, data mining, data structures and algorithm analysis.

**Nasrullah Memon holds** a PhD in Intelligence Security and Informatics (Investigative Data Mining) from Aalborg University Denmark in 2007. He has also Masters degrees in Applied Mathematics and Software Development from University of Sindh and University of Huddersfield. His research interests include machine learning, natural language processing, information retrieval, information extraction, open source intelligence and investigative data mining. He has published more than 90 research articles in journals/ conferences/ book chapters of national and international repute.